\documentstyle[preprint,aps,psfig,epsf]{revtex}

\newcommand{\ave}[1]{\mbox{$\langle #1 \rangle$}}

\newcommand{\beq}{\begin{equation}}
\newcommand{\eeq}{\end{equation}}
\newcommand{\beqa}{\begin{eqnarray}}
\newcommand{\eeqa}{\end{eqnarray}}
\newcommand{\bmath}{\begin{mathletters}}
\newcommand{\emath}{\end{mathletters}}

\newcommand{\Istyle}{\mbox{$ {\cal I}$}}

\newcommand{\helium}{\mbox{$^3{\rm He\!-\!B}\;$}}
\newcommand{\bc}{\mbox{{\em bc }}}
\begin{document}
\draft
\preprint{ICTP-SISSA} 
%\advance\textheight by 0.2in
%\twocolumn[\hsize\textwidth\columnwidth\hsize\csname@twocolumnfalse%
%\endcsname

\title{Phase oscillations in superfluid \helium weak links}

\author{A.~Smerzi$^1$, S.~Raghavan$^{2,3}$, S.~Fantoni$^{1,2}$,
and S.~R.~Shenoy$^2$}
\address{$^1$ 
 Istituto Nazionale di Fisica
della Materia and 
International School for
 Advanced Studies,
I-34014, Trieste, Italy \\
$^2$ Abdus Salam International Center for Theoretical Physics,
 I-34100, Trieste, Italy \\
$^3$ Rochester Theory Center for Optical Science \& Engineering, Department
of Physics \& Astronomy,\\
University of Rochester, Rochester NY 14627}

\date{\today}
\maketitle
\begin{abstract}
Oscillations in quantum phase about a mean value of $\pi$, 
observed across micropores connecting two \helium baths, 
are explained in a Ginzburg-Landau phenomenology.
The dynamics arises from the Josephson phase relation,the interbath 
continuity equation, and helium boundary conditions.
The pores are shown to act as Josephson tunnel junctions,
and the dynamic variables are the inter bath phase difference
and fractional difference in superfluid density at micropores. 
The system maps onto a non-rigid, momentum-shortened
pendulum, with inverted-orientation oscillations about a vertical tilt
angle $\phi = \pi$, and other modes are predicted.
\end{abstract}
%\section{}
%\twocolumn
%]

The search for Josephson-like weak-link effects  in
superfluids has a
long history \cite{tilley90,anderson66,avenel85,varoquaux90,%
pereverzev97,backhaus98,simmonds98}. In order to see
Josephson-like phenomena, 
the characteristic length-scale of the weak-link
should be comparable to the 
temperature-dependent 
coherence length $\xi(T)$. Thus 
a candidate superfluid is \helium, with a 
relatively large zero-temperature coherence length $\xi(0) \sim 65$nm
\cite{vollhardt90}. 
In fact the oscillatory displacement of flexible membranes 
walling off \helium baths connected  by micropores, induces a 
Josephson-like, periodic current-phase relation  
 \cite{avenel85,varoquaux90,pereverzev97,backhaus98,simmonds98}. 
Recently, Davis, Packard, and collaborators have observed a remarkable
phenomenon:
metastable oscillations of \helium with an average 
phase-difference of $\pi$ across the weak link \cite{backhaus98}. 
The explanation of these, and other
complex tunneling oscillations is of considerable general interest, since
they provide examples of novel
macroscopic quantum effects.

 Bose-Einstein condensate (BEC) tunneling of neutral atoms between
double-well
traps has been predicted \cite{sfgs-rsfs} to support $\pi$-state
oscillations, and other modes. The coupling between the
$N_{1,2}$ condensate atoms in wells 1,2, with phase difference $\phi$
is $-\sqrt{N_1N_2}$cos$\phi \sim -( 1-z^2)^{1/2}$cos$\phi$ where 
$z =(N_1-N_2)/(N_1+N_2)$
is the fractional population imbalance.
The coupling energy corresponds to \cite{notea} a  {\it non-rigid} pendulum
\cite{sfgs-rsfs,marino} of
tilt angle $\phi$, momentum $p_{\phi} = z$ and length $(1-
{p_{\phi}}^2)^{1/2}$. 
The non-rigid pendulum is shortest when moving fastest, and can thus
support inverted-orientation oscillations about $\phi = \pi$ (concave
downwards). The BEC mapping onto a non-rigid pendulum dynamics is the
generic consequence of wave function phase factors and the two-state
nature of the tunneling: properties that it shares with the
\helium system. However,the BEC dynamics cannot be naively taken over
\cite{noteb}: features of the
constrained helium wavefunction in a pore geometry, must play a role.

                  In this Letter we show that the helium wave function
boundary conditions in a pore geometry, in conjunction with the
Josephson phase relation and the continuity equation, determine 
the \helium  
tunneling dynamics, that is indeed similar to non-rigid pendulum
dynamics, but with a
(momentum-dependent)
torsion-bar \cite{fulton} due to hydrodynamic flows outside the pore
\cite{varoquaux90}.
Thus $\it {dynamical}$ $\pi$-states of
temperature-dependent frequency
can occur, with a rich variety of other modes.

                   The wavefunction is depressed by the boundary
conditions (\bc) inside, and just outside
a pore length, that is shown to act as a tunneling barrier, for
sufficiently
narrow pores. Displacements of the
flexible membrane wall  are proportional to
a fractional shift $z$ of the depressed wavefunction, that
acts as an oscillating
`Josephson piston': it  conveys pairs
through the tunneling region
at a rate ($\sim \dot{z}$) determined by the (sine of) the phase
difference
$\phi_J$ across the pore. 
We now derive the  (quasiequilibrium) \helium dynamics.

{\it Free Energy}: The \helium order parameter \cite{vollhardt90} can be
written as 
$\Phi_{\mu ,i}({\vec r}) = \Psi ({\vec r}) B_{\mu , i} (${${\vec
n}({\vec r})$}$)$,where
$\Psi ({\vec r})=|\Psi ({\vec r})| e^{i\phi({\vec r})}$
and $|\Psi({\vec r})|$ is proportional to the  gap, varying on a
coherence length  $\xi(T)$. Here 
${\bf B}({\vec n}({\vec r}))$ is a tensorial factor,with `${\vec n}({\vec
r})$'  representing axis rotations,varying over a
textural healing length  $l_{tex}, \geq 5 \mu m$
\cite{vollhardt90}.The
Ginzburg-Landau
(GL) free energy is
a scalar, with traces over the tensorial products: $F(\{{\bf \Phi}
({\vec r}),\nabla
{\bf \Phi} ({\vec r})\}) = F_0(\{{\bf \Phi} ({\vec r})\}) +F_{grad} (\{\nabla
{\bf \Phi}({\vec r})\})$ \cite{vollhardt90}. 
 Unlike the
superconductor \bc  of vanishing normal gradients of the gap,
we must use here `helium' \bc, with the order parameter 
vanishing at the walls, $\Phi_{\mu ,i} ({\vec r}) \rightarrow 0$ .
This \bc, supported by 
experiment, has been used elsewhere
\cite{ullah89} ,and is particularly
relevant
in the system regime \cite{pereverzev97,backhaus98,simmonds98} of pore
scales $\sim \xi, \ll l_{tex}$.
The \bc is enforced by $\Psi \rightarrow 0$ over length scales $\sim
\xi$, and the contributions to $F_{grad}$ from $|\nabla \Psi|$ 
contributions will dominate those from $|\nabla B| \sim |\nabla n |$
\cite{notec}.
The ${\vec n}({\vec r})$ textural variables can be set equal to their 
fixed,bath value, 
\cite{vollhardt90}.
%thuneberg88}.
Thus $F_{grad} \rightarrow F_{grad} (\{\nabla \Psi(r)\})$,
and the effective  free energy describing
the pore region is 
\beq
F =  \int d^3 r \left[{{\hbar}^2\over {2 m^{*}}} \left|{{\nabla
{\Psi(\vec r)}}
} \right|^2
+ a \epsilon|{\Psi}|^2 + {b \over 2} |{\Psi}|^4 \right] 
\label{eq:eq1}
\eeq
where $a,b,m^{*}$ are GL coefficients (constants in space and time)
absorbing tensorial-factor traces of order unity; and $\epsilon =
(T/T_c-1)$, where $T_c$ is the transition temperature.

           The experimental geometry \cite{pereverzev97} is a superfluid
of total mass density
$\rho$ in a pillbox of cross-section $A$, and length $L$,
closed at one end by a flexible membrane (of spring constant $C$)
and at the other end by a perforated wall with $N_J =65 \times 65$
pores of geometric cross-section $A_{J0}$ and length $L_{J0}$. The pillbox
bath is immersed
in a larger superfluid bath.See Fig 1.We consider a single pore.
The $x$ coordinate  is along
the pore axis, and the system cross-section $A(x)$ varies between $A$ and
$A_{J0}$.

               To obtain an effectively 1D model, we  average
over the transverse cross-section of the system, and
scale in the equilibrium  wave function 
$|\Psi_0|$ =$(a |\epsilon | /b)^{1/2}$ where the effective superfluid mass
density is 
$\rho_s (T) = m^{*} |\Psi_0|^2 \sim |\epsilon|$. Thus the  1D
gap
wavefunction is $\psi(x)$ = [$\Psi ({\vec r})$]/$|\Psi_0|$,where [...]
$\equiv \int d^2r (...)/A(x)$. With ${\xi (T)}^2 = {\hbar}^2 /(2m^{*} a
 |\epsilon|)$, the  free energy from Eq.(1) is: 
\beq
F \simeq \epsilon_0 \int dx A(x) \left[
\xi^2 |{\partial}_x {\psi} |^2+
(U-1)|{\psi}|^2 +
{
1\over 2}|{\psi}|^4 \right]
\label{eq:eq2}
\eeq
where the  energy density is $\epsilon_0 \equiv |\Psi_0|^2 a |\epsilon| 
\sim |\epsilon|^2$.
% and $\beta (x) \equiv
%[|\Psi (\vec r)|^4]/|[\Psi (\vec r)]|^4$,that
%is unity in the bulk.  
The averaged transverse gradients $U(x)=[|{\partial}_y \Psi ({\vec r})|^2
+
|{\partial}_z 
\Psi ({\vec r})|^2]/|[\Psi({\vec r})]|^2$
are $\sim {\xi }^2/A_{J0}$ in the pore region, and negligible,
$\sim {\xi }^2/A$ in the bulk. Thus $U(x)$ is an effective energy
barrier,
arising from the \bc. We set $U(x)= {\gamma}^2{\xi }^2
/A_{J0}$ in an effective barrier length $|x| < {L_J}/2$, and zero outside
it.The barrier length $L_J$ is greater than the
geometric pore length
$L_{J0}$, as the wave function depression (and large transverse gradients)
will persist by continuity, for a small distance or `overhang', outside the
pore openings,of size  $\delta =(L_J
-L_{J0})/2$ as shown in Fig 1. We treat 
$\gamma$, $> 1$, as a fitting parameter, and consider $e^{-K L_J} \ll 1$
as small.

{\it Wavefunctions}: The barrier region solutions of the (linearized)
 GL equation derived from Eq.~(\ref{eq:eq2})
 are $e^{\pm K x}$, with a real  decay
wavevector 
$K \equiv \sqrt{(\gamma^2 / A_{J0}) - \xi^{-2}}$, for $\sqrt {A_{J0}}
< \gamma \xi$.( An estimate, for square pores, is $\gamma = \sqrt{2}
\pi$.) 
With a phase difference $\phi_J \equiv \phi(x=L_J/2) - \phi(x= -L_J/2)$
across the junction,the (dimensionless) pore wavefunctions are in
 a form \cite{varoquaux90} generalized to 
\beqa
{\psi}(x) = [&a_-& e^{- i \phi_J/2} {\sinh K ((L_J/2) -x)} + \nonumber \\ 
&a_+& e^{i \phi_J/2} {\sinh K ((L_J/2) + x)}]/ {\sinh K L_J} ,  
\label{eq:eq3}
\eeqa
where coefficients $a_{\pm}$ here match on to 
bath wave functions, rather than  plane waves \cite{varoquaux90}.
Substitution into Eq.~(\ref{eq:eq2}) yields a leading-order, $|x| \leq
L_J/2$ contribution 
\beq
F_J = -E_J \sqrt{(a_+^2 a_-^2)/a_0^4} \cos \phi_J,
\label{eq:eq4}
\eeq
neglecting  $O (e^{-2KL_J}\cos {2 \phi_J})$ corrections.
The Josephson energy  is
$E_J \equiv {2 A_{J0} K {\xi}^2 \epsilon_0 a_0^2 f(T)}/
{ \sinh {K L_J}}$, and
$f(T)$ is a possible renormalization factor discussed 
later. We now estimate $a_{\pm}$ 
amplitudes 
and their dependence on interbath  pair transfers. 

Equilibrium wave-functions deep in the baths $1,2$
are rigid,
as bulk \helium is incompressible. Solutions of the nonlinear GL equation 
from Eq.(1) 
are hyperbolic tangents, 
flat in the bulk and falling to zero with slope 
$1/ \sqrt{2} \xi$ at walls \cite{tilley90}. We take the averaged bath
wavefunction as 
$\psi_{1,2}(x) = e^{\mp i \phi_J /2} g(\mp x - L_{J0}/2)$, where 
$g(x) = 
\tanh(x/\sqrt{2} \xi)$. At $x = \mp L_J/2$, 
continuity with Eq.~(\ref{eq:eq3}) yields equal equilibrium
pore-amplitudes 
$a_- = a_+ \equiv  a_0 $, and $a_0 =g_0$, where 
the bath wavefunction at the barrier edge is $g_0 \equiv g(\delta)$.
See Fig. 1.
Slope-matching determines the (weakly $T$-dependent) overhang $\delta$ 
as $\sinh(\sqrt{2} \delta / \xi) 
= \sqrt{2} / K \xi$ coth$((K( L_{J0} + 2\delta)/2)$. For long narrow pores,
 $g_0 \sim \delta/\sqrt{2 
} \xi$, and  the overhang $\delta \simeq
1/K \sim A_{J0}^{1/2}$ vanishes, as pores close, $A_{J0} \rightarrow 0$.

Slowly oscillating and small ($X(t) \ll \xi$) 
displacements  of the far-off flexible membrane  will be transmitted
by the 
rigid bulk, to induce shifts by
$\eta (\propto X)$ of the  wavefunction near the pores. 
Since $\eta$ 
will be
related to the tunneling through the pores, we term the movable region of 
length $L_J$ in Fig. 1, a 'Josephson piston'. The bulk wavefunctions 
at $x = \mp L_J / 2 + \eta$ then
match onto the (shifted) piston Eq.~(\ref{eq:eq3}),
yielding $a_{\pm}$.
For small $\eta$, defining $z \equiv 2 \eta / \delta$
as a fractional piston-overhang shift, 
we can expand as 
$\sqrt{a^2_+ a^2_-} / a_0^2 \sim \sqrt{1 - z^2}$
to obtain  
Eq.~(\ref{eq:eq4}) as $F_J=F_J(\phi,z)$.(Additional
wavefunction stiffness terms, $F_J \sim  +z^2$
make $z = 0$ the piston rest position; these are dominated by
membrane stiffness terms, given later).

The shift of the interfaces between the
Josephson region and the bath
to  $x = -{1\over 2}L_J + \eta,~x = {1 \over 2}
L_J + \eta$ means that there are increases/decreases
in the number of atoms in the two tunneling overhang regions.
Note that  
$z$ can be interpreted as a ``BEC-like'' fractional
population imbalance \cite{sfgs-rsfs} of the overhang volume 
$2  A_{J0} \delta$, and is {\it not} a fraction of the total Cooper pairs
\cite{noteb}.

 Clearly, the 
incremental number change $dn_{1,2}$in each bath
will be the number density at the interface times the incremental shift 
$d \eta$. The change in tunneling population 
$d n = {1 \over 2} (d n_1 - d n_2)$ from the overhangs on the two sides is
then
\cite{raghavan}
$d n =
- (\rho A_{J0} g_0^2  \delta /2 m_3) dz$ where $m_3$ is the helium
atom mass.
The piston displacements induce a reactive velocity component
\cite{raghavan} 
of the membrane $\dot{X} \propto {1\over 2} (g_0^2 \delta) \dot{z}$ 
proportional to the tunneling current. 

{\it Current Equation:}The tunneling current (piston velocity) equation
for
$\dot{z}$  is provided by the continuity equation 
$\dot{\rho}(r) = \vec{\nabla}\cdot
\vec{ J}(r)$ where $\vec J$ is the total mass current density.  
Integrating over baths $1,2$ (outside the piston regions) 
and using Gauss's theorem, 
$m_3 \dot{n} = {1 \over 2}A_{J0}( [J_x(x = -L_J/2)] +
[J_x(x = L_J /2)])$. 
Assuming  no normal 
fluid flows through the
narrow pores, and with superfluid mass-current 
density
 $[J_x(x)] = \rho_s {\hbar \over m^{\ast}} Im (\psi^{\star} \partial \psi/ 
\partial x)$,  Eq.(\ref{eq:eq3}) yields:
\beq
\frac{\hbar}{2 K_J} \frac{dz}{dt} =  -\sqrt{1 - z^2} \sin \phi_J =
-{{\partial H}\over {\partial \phi_J}}.
\label{eq:eq5}
\eeq
where $H$ and $K_J$ are defined below.The physical picture is  of phases
at $N_J$ pore openings in each bath locked
together,and with a common difference between baths,so that the  $4000$
Josephson pistons vibrate in unison.

{\it Phase Equation:} The Josephson relation for $\phi = \phi_2 -\phi_1$,
the total
interbath phase
difference,  is $\dot{\phi}=\Delta \mu/\hbar$
and the chemical potential difference from the tunneling transfer
of $n$ atoms will contribute as 
$\Delta \mu \simeq -{{\partial F_J} / {\partial n}}$.(The bulk
contributions 
from the incompressible, constant-density fluid in the baths, clearly 
cancel out of 
the  difference  $\Delta \mu$.)
Work done by a fixed
\cite{simmonds98} external 
pressure $P_{ext}$, and by the Josephson piston, pushing against the membrane,
of spring constant $C$, can also be included 
\cite{raghavan}. 
Thus 
\bmath
\label{eq:eq6}
\beqa
\frac{\hbar}{2 K_J} \frac{d \phi}{dt}
&=&  \Lambda z +{P_{ext}\over {P_J}} + 
{z  \over {\sqrt{1-z^2}}} \cos
{\phi_J}= {{\partial H}\over {\partial z}}; \\
H &=& {1 \over 2} \Lambda z^2 + {P_{ext}z \over P_J} - 
\sqrt{1 - z^2} \cos \phi_J,
\eeqa
\emath
where $P_J \equiv \rho K_J/m_3$ is a pressure scale. The
frequency scale is $2 K_J/\hbar  \equiv (\rho_s / \rho) ({2 \hbar K}/{m^{*}
\delta}){f(T) / {\sinh( K L_J)}}  \sim |\epsilon| f(T)$;
and  
$\Lambda \equiv ({C g_0^2 \delta })/( {2 A  P_J}) \sim C/f(T)$
is a (dimensionless) membrane stiffness parameter. 

Note that $\phi$ and $z$ are not canonically conjugate,as $\phi \neq
\phi_J$
due to hydrodynamic effects \cite{varoquaux90}. 
The hydrodynamic bath-pore superflow $I = (A_h/A_{J0})A_{J0}  
\rho_s (\hbar/ m^{*}) (\phi_h/ L_h)$ 
is driven by a phase gradient
$\phi_h / L_h$  over a hydrodynamic length $L_h$ in each bath. 
(Here $A_h/A_{J0}$ is a bath-pore `transmission probability').        
The same current $I = I_J \sqrt{1 - z^2} \sin {\phi_J}$, tunnels through
the pore, where  the Josephson critical (mass) current  is
$I_J = E_J 2 m_3/\hbar$ 
$\sim |\epsilon |^2 f(T)$.
With total phase change $\phi = 2 \phi_h + \phi_J$, 
we have a nonlinear relation  
\cite{varoquaux90} generalized to $\phi = \phi(\phi_J, z)$: 
\beq
\phi = \phi_J + \alpha \sqrt{1 - z^2} \sin \phi_J, 
\label{eq:eq7}
\eeq
where 
$\alpha \equiv (2 {A_{J0} \over A_h}) ({K  L_h g_0^2 f(T)})/ 
{\sinh (K L_J) } \sim |\epsilon| f(T)$.
This corresponds to a torsion-bar as in the inductive SJJ case
\cite{notea,fulton}, but
now,momentum-dependent.

The temperature-dependence of $\Lambda (T)$ is 
essentially  that of $f(T)$.
Experimentally,the critical current  goes as $\sim (1-T/T_{ca})^2$,
appearing to vanish before transition, at a $T_{ca}= 0.91 T_c$
\cite{backhaus98}. Since
$I_J \sim {\epsilon}^2 f(T)$, this `experimental' form $f(T) =
(1-T/T_{ca})^2/{\epsilon}^2$, could be
used.Or,
$f(T)$ could be attributed to 
thermal phase fluctuations  
\cite{granato}.We estimate it, in this picture \cite{noted}.

Eq.~(\ref{eq:eq5},\ref{eq:eq6},\ref{eq:eq7}) constitute the
model equations for the \helium system, in terms  of the mutually linked 
dynamics of the fractional 
Josephson piston shift $\eta / \delta \equiv z /2$
and the  phase difference $\phi_J$ across it.
They can be  solved in terms of elliptic functions \cite{raghavan}
just as in the $\alpha =0$ `BEC-like' case
\cite{sfgs-rsfs}.They correspond to a non-rigid pendulum of tilt 
angle $\phi_J$,and  angular momentum $z$,with a momentum-dependent torsion
bar. The torsion-bar \cite{fulton} induces hysteretic effects for 
$\alpha \geq 1$.We focus on  $\alpha <  1$ where the modes are
essentially as for  $\alpha = 0$ , with the equations predicting 
five distinct undriven modes (instead of the two of the rigid pendulum).
The modes can be visualized by plotting \cite{marino} the locus of the
pendulum
coordinates $(\sqrt {1-z^2} \sin \phi_J, -\sqrt {1-z^2} \cos \phi_J)$.             
     With a dc drive $P_{ext}$ there is an ac Josephson
oscillation at a frequency $\omega_{ac} = 2m_3 P_{ext}/{\rho \hbar}$;
with added resonant ac drives, there are Shapiro-like dc current spikes
\cite{sfgs-rsfs,raghavan}.
`Zero-state' oscillations of the pendulum can occur, with time averaged
$\ave {\phi} = 0 = \ave {z}$.The pendulum, once excited, can rotate
freely, with a running phase, and a `self-trapped'
piston-shift/membrane-displacement $\sim \ave {z} \neq 0$.
               
              Since the non-rigid  pendulum is shortest, when moving
fastest,it can  execute small `inverted'
oscillations  about an average value of
$\ave {\phi} = \pi$. These $\Lambda \lesssim 1$, 
`$\pi$-state oscillations', with $\ave {z} = 0$, and are {\it dynamically}
metastable, with the momentum-shortened pendulum `digging a well for
itself'
by its motion. Finally, there are two types of $\Lambda >1$, `$\pi$-state
rotations'
with self-trapped  $\ave {z} \neq 0$ and $\ave {\phi} = \pi$,(on either 
side of a fixed point $\phi =\pi,z=\sqrt {1- {\Lambda}^{-2}}$).These
correspond to the non-rigid
pendulum executing closed-loop trajectories,floating above the point of
support \cite{sfgs-rsfs,marino}. Damping of $\pi$-state oscillations
through the non-rigid pendulum
velocity $\dot {\phi}$ , or the momentum $z$, behave quite differently.
While the former \cite{marino} drives the system to rest at $\phi_J = 0$,
the latter \cite{raghavan} damps it to $\phi_J =\pi$.

We use values \cite{backhaus98} 
$L_{J0} = 0.5~10^{-5}$ cm, $A_{J0} = 10^{-10}$ cm$^2$,
$\rho = 0.08$ g/cm$^3$, $ m^{*} \simeq 2m_3 = 10^{-23}$ g, 
$N_J = 4225,
~\xi(0) = 65$ nm,
$A = 0.7$ cm$^2$, $C = 10^6$ dynes/cm, $T_c = 0.91$ mK,
and a purely
illustrative parameter $\gamma \simeq 23$ 
. Then
%$K \simeq 2.6~10^6$ cm$^{-1} = \delta^{-1}$, and
$\delta  \simeq 90 {\rm \AA}$ and
$2 K_J(0) / \hbar \simeq 400$ Hz, 
$(I_J(0)/m_3) \simeq 3.2 \times 10^6$ atoms/sec, corresponding to a Josephson
energy per pore $E_J(0) \simeq 0.02~k_B T_c$.
 We find $\Lambda(0) \simeq 0.9;
P_J \simeq 0.35$ mPa; and $\alpha(0) \simeq 0.5$. 
\par

       The results of our model are shown in Fig.2. The dimensionless
membrane displacement versus 
dimensionless time ($t 2 K_J/{\hbar} \rightarrow t$) is shown  for
zero- and $\pi$-states. The insets show the temperature-dependent 
frequencies and critical currents. These include:
i) zero-state oscillations , with harmonic frequencies 
$\omega_{0}(T) = [(1 + \Lambda)/ (1 + \alpha)]^{1 \over 2}
(2K_J/ \hbar)$
that vary from kHz,   to zero at $T_c$. 
ii) $\pi$-state
oscillations, with  smaller amplitudes and frequencies 
$\omega_{\pi} (T) = [(1 - \Lambda) / (1  - \alpha )]^{1/2} (2
K_J/\hbar)  < \omega_0 (T)$, 
that vanish on warming, when $\Lambda(T)$ crosses unity, well before $T_c$.
iii) A critical current \cite{heliumnota2} ($\lesssim 20$ picogm/s) goes as
${\epsilon}^2$
,appearing to vanish before $T_c$. iv) With a phase
angle defined as the time-integrated membrane displacement \cite{backhaus98}, the
current-phase relations (not shown) have same-sign, positive slopes 
in the zero- and $\pi$-states.
            
Our model clearly reproduces  experimental features of the
magnitudes/temperature
dependences,of the oscillations, their frequencies, and currents
\cite{pereverzev97,backhaus98}.
It also predicts  self-trapped $\pi$-states with
${\omega_{\pi}}^{tr} = [(\Lambda^2 -1)/(1 - \alpha / \Lambda)]^{1 \over 2}
(2 K_J / \hbar)$,  
that have not yet been reported.

Further work could include  
experimental investigations of predicted
intermode
transitions with frequencies dipping to zero, as
$\Lambda(T)$ or $z(0),\phi (0)$ are varied through critical values
\cite{sfgs-rsfs,raghavan}; and a search for self-trapped $\pi$-states.

$\pi$ states have been related to a 
static metastable minimum 
induced by spin-textures in the
baths \cite{Yip99}.
This could dress our dynamic $\pi$ states from oscillations of the 
density depression at pores,
producing  fine structure in the oscillation modes. In fact, 
bistability between two types of $\pi$ oscillations has been reported
\cite{Marchenko99}.

In conclusion, we have modelled the dynamics of \helium baths closed by 
flexible walls, and connected by micropores. The dynamic equations arise
from helium boundary conditions,the continuity equation ,and  the
Josephson phase relation. The dynamical variables are the phase
difference, and the shift of the pore wave-function overhang, beyond 
the geometric pore length. The system maps on to a non-rigid 
momentum-shortened pendulum
dynamics.
The $\pi-$phase oscillations observed in \helium 
have features similar to those predicted by the model.
Other tunneling modes are predicted.

It is a pleasure to thank J.C.Davies, R.E.Packard, 
E.Granato,M. Mehta, R.Simmonds and S. Vitale 
for useful conversations. We thank R.E. Packard for providing 
unpublished data. This work was partly supported by the
Cofinanziamento MURST and by the NSF Grant PHY94-15583.

%\bibliography{/home/srirag/tex/bibtexbase/thesbib,%

\begin{thebibliography}{10}
\bibitem{noted}
{A  gaussian `spin-wave' average of phase factors   
  with a  coupling energy proportional to
  $N_J E_J 
\sim |\epsilon|^2$ yields an estimate of
$f(T) \sim
  e^{-(T/T_c)(\epsilon_1/\epsilon)^2}$, and  typically,
$\epsilon_1   \sim 0.2$. The singular flattening of this $f(T)$ at $T =
T_c$  makes it  seem to vanish at a `$T_{ca}$'$ < T_c$. 
The  stiffness $\Lambda(T) = \Lambda(0) / f(T)$ diverges at $T_c$.}
\bibitem{heliumnota2}
{If the measured signal in the $\dot X$ membrane
velocity is presented as a mass current 
$\Istyle \equiv \rho A \dot{X} / N_J$ \cite{backhaus98}, 
then a related characteristic current is 
$\Istyle_c = (A/N_J A_J) I_J$.}
\end{thebibliography}
%/home/srirag/tex/bibtexbase/bec,%
%/home/srirag/tex/bibtexbase/mesoscopic,%
%/home/srirag/tex/bibtexbase/helium}

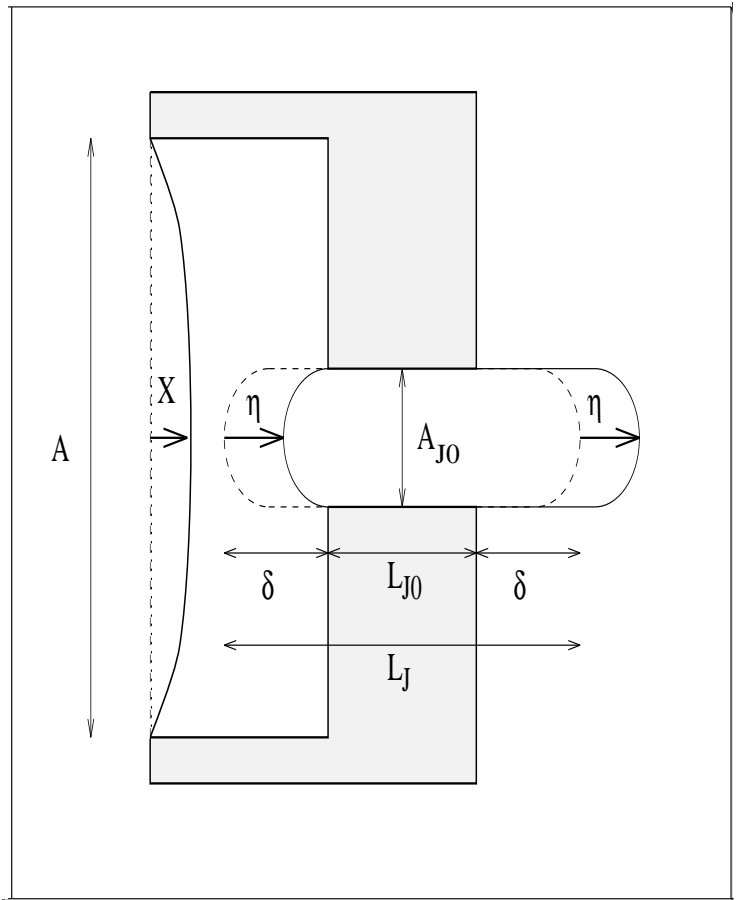
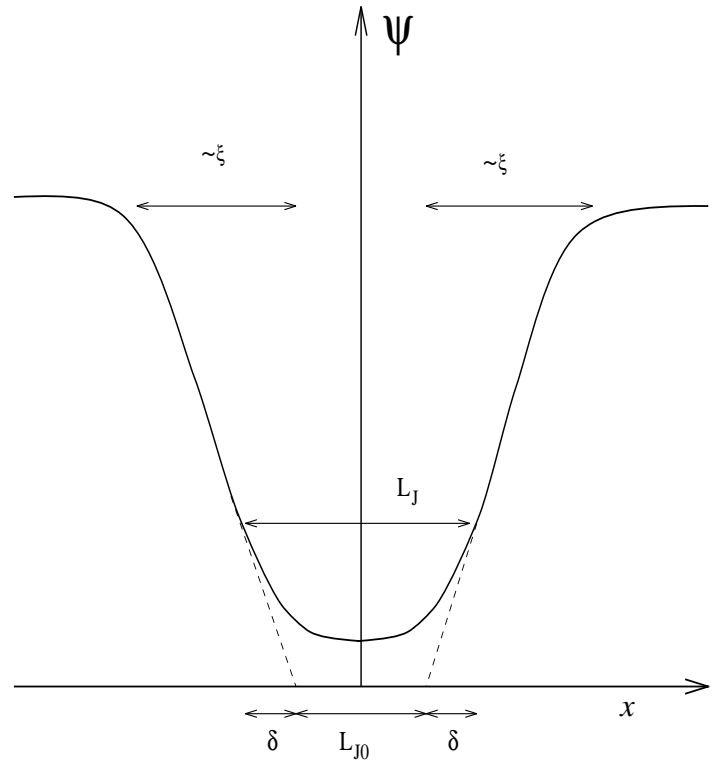
\begin{figure}
\caption{Schematic diagrams (not to scale) of : a)Effective pore/bath
geometry,
showing
geometric pore
length $L_{J0}$ and pore (bath) area $A_{J0}$ ($A$).Dashed line shows the
rest position of the  `Josephson piston' of 
length $L_J= L_{J0} + 2 \delta$.The piston shift is $\eta$ and the
membrane displacement is $X (\propto \eta)$.  ; b)Equilibrium wavefunction
$\psi(x)$ versus $x$. } 
\label{fig:fig1}
\end{figure}

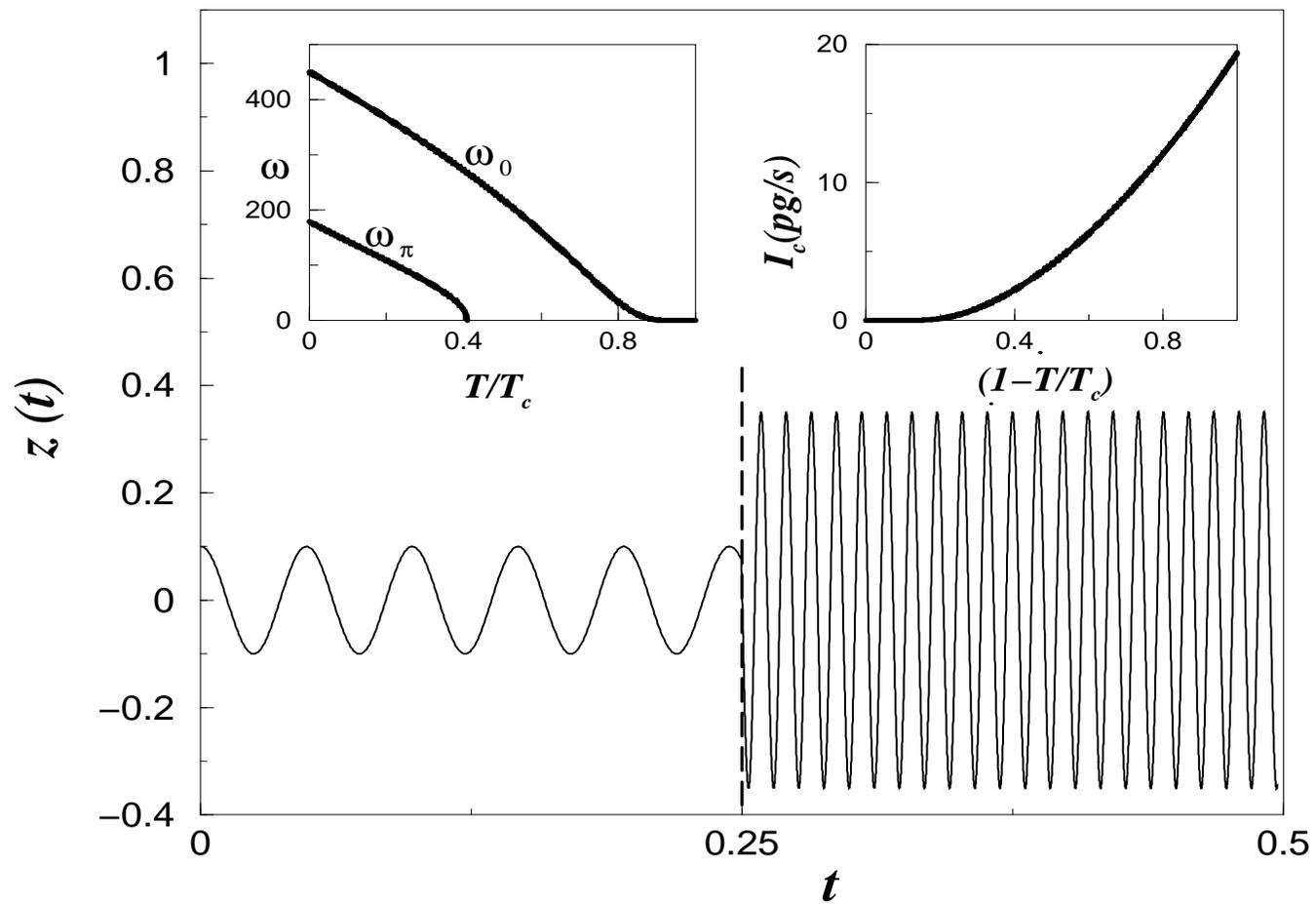
\begin{figure}
\caption{Dimensionless $z(t)$ (proportional to membrane displacement)
versus dimensionless time 
for $|\epsilon| = 1$,
and parameters as in text.
$\pi$-state oscillations (early times); are 
kicked out to zero phase 
%$\ave{z}=0, \ave{\phi} =0$ 
oscillations
 (later times).Left inset:Zero-state ($\pi$-state)
angular frequencies $\omega_0(T)~  (\omega_\pi(T))$ defined in text,
versus temperature.
Right inset: Characteristic current 
$\Istyle_c
% = (A/ N_J A_J) I_J$ (in picograms/sec),proportional to 
\propto I_J(T)$ the Josephson critical  current,  
as defined in text, %and \cite{heliumnota2}
 versus temperature.}
\label{fig:fig2}
\end{figure}

%\begin{figure}
%\caption{Current-phase plot $I(\phi) / I_{max}$ versus
%$\phi$ from Eq.(8), $\alpha= 0.5$ and illustrative parameters as in the
%text.
%$\pi$-state
%branches are seen, around $\phi = \pi, 3 \pi...$.}
%\label{fig:fig3}
%\end{figure}

\end{document}